\documentclass[showpacs,prd]{revtex4}
\usepackage{epsfig}
\usepackage{graphicx}
\usepackage{dcolumn}
\usepackage{amsmath}
\usepackage{latexsym}

\begin{document}

%%%%%%%%%%%%%%%%%%%%
\title{Path integral action and Chern-Simons quantum mechanics in noncommutative plane}  
%\date{\today}
\author{Sunandan Gangopadhyay$^{a,b,c}$\footnote{e-mail:sunandan.gangopadhyay@gmail.com}, 
Frederik G Scholtz $^{a,d}$\footnote{e-mail:
fgs@sun.ac.za}}
\affiliation{$^a$National Institute for Theoretical Physics (NITheP), 
Stellenbosch 7602, South Africa\\
$^b$Department of Physics, West Bengal State University, 
Barasat, Kolkata 700126, India\\
$^c$Inter University Centre for Astronomy and Astrophysics (IUCAA), Pune 411007, India\\
$^d$Institute of Theoretical Physics, 
Stellenbosch University, Stellenbosch 7602, South Africa}

%%%%%%%%%%%%%%%%
\begin{abstract}
\noindent In this paper, the connection between the path integral representation of propagators in the coherent 
state basis with additional degrees of freedom \cite{rohwer} and the one without any such degrees of freedom 
\cite{sgfgs} is established. We further demonstrate that the path integral formalism developed in the noncommutative plane
using the coherent state basis leads to a quantum mechanics involving a Chern-Simons term in momentum which is of noncommutative origin. The origin of this term from the Bopp-shift point of view is also investigated. A relativistic generalization of the action derived from the path integral framework is then proposed. Finally, we construct a map from the commutative quantum Hall system to a particle in a noncommutative plane moving in a magnetic field. The value of the noncommutative parameter from this map is then computed and is found to agree with previous results.

%%%%%%%%%%%%%%%%

\end{abstract}

\maketitle

\noindent In the last decade, quantum mechanics on the noncommutative plane \cite{duval}-\cite{sgthesis}
and noncommutative quantum field theory \cite{witten}-\cite{carroll} 
has been one of the intense areas of research owing to a continuous search for
a consistent theory of quantum gravity. A key feature of these theories (like other theories of quantum gravity
such as string theory) is that they admit a description in terms of extended objects, as first pointed out in \cite{suss} where a mapping to electric dipoles moving in a magnetic field was established. However, to get a deeper
insight of the notion of physical extent and structure, one has to pay careful attention to the formal and 
interpretational aspects of noncommutative quantum mechanics. These aspects were formulated in \cite{fgs}
by viewing noncommutative quantum mechanics as a quantum system represented
on the space of Hilbert-Schmidt operators acting on noncommutative configuration space.
A path integral formulation was also developed, which clearly demonstrated the nonlocal nature of noncommutative 
theories. Based on these investigations, it was shown in \cite{rohwer} that there exists two equivalent pictures in position representation: a constrained local description in position containing additional degrees of freedom and an unconstrained
nonlocal description in position without any other degrees of freedom. Both these descriptions clearly point towards
the notion of extended objects.

In this paper, we proceed to derive the path integral representation and the action for a particle using coherent states
containing additional degrees of freedom. 
We follow the standard approach of introducing momentum completeness relations
\cite{sgfgs, sgfgs1} to derive this representation.
The advantage of doing this is that it leads to a phase-space representation of the path integral.
We also point out that this action has a structure very similar to the action involving an auxiliary field considered in
\cite{sgfgs}. As a matter of consistency, we quantize this system by noting that it is a second class constrained system
and recover the noncommutative Heisenberg algebra.
The connection between this path integral picture
with the one formulated earlier in \cite{sgfgs} without any additional degrees of freedom is then established. 
The free particle propagator involving the coherent states with additional degrees of freedom is computed next from which the propagator in the coherent state basis (without any additional degrees of freedom) is recovered.
We then observe that the path integral formulation on the noncommutative
plane in the coherent state basis naturally leads to a noncommutative
Chern-Simons quantum mechanics. This is shown by first obtaining the
phase-space representation of the path integral in the coherent state basis $|z)$. 
Interestingly, one finds that the action contains a Chern-Simons
term in momentum thereby making it similar in form with the path integral in $|z; v)$ coherent state basis.
Indeed, such a theory was considered earlier in \cite{luk}, \cite{jianjing} where the Chern-Simons term in momentum was put in by hand in order to make a transition from a commutative theory to a noncommutative theory.
However, the path integral formulation presents a clear justification of introducing 
such a term (of noncommutative origin) in the Lagrangian.
This observation is completely new and has been missing from the existing literature.
It also demonstrates that there is a subtle connection between the origin of the Chern-Simons term in momentum and the Voros star product. We also obtain the phase-space representation of the path integral for a particle in the noncommutative plane in the presence of a magnetic field. This leads to an action for a particle in a magnetic field augmented by the Chern-Simons term in momentum. We then try to understand the origin of this term from the Bopp-shift view point. It is observed
that replacing the position coordinate in the first order form of the action by the Bopp-shifted position coordinate gives rise
to this term. An action for a relativistic particle in the noncommutative plane is then written down by
a straight forward generalization of the non-relativistic action derived from the path integral framework.
Finally, we consider the problem of a quantum Hall system in commutative space. We show that the problem can be mapped
to the problem of a particle in a noncommutative plane in the presence of a magnetic field. The value
of the noncommutative parameter can be computed explicitly and is found to be in agreement with the result for the commutator of the relative coordinates (projected to the lowest Landau level) in a magnetic field in the presence of a harmonic oscillator potential computed in \cite{sggov}.

%%%%%%%%%%%%%%%%%%%%%%%%%%%%%%%%%%%%%%%%%%%%
To begin our discussion, we present a brief review of the formalism of 
noncommutative quantum mechanics developed recently in \cite{fgs}. It was suggested in these papers that one can give
precise meaning to the concepts of the classical configuration space and the Hilbert space
of a noncommutative quantum system. The first step is to define 
classical configuration space. In two dimensions, 
the coordinates of noncommutative configuration space satisfy the commutation relation 
\begin{equation}
[\hat{x}, \hat{y}] = i\theta
\label{1}
\end{equation} 
for a constant $\theta$ that we can take, without loss of generality, to be positive. The annihilation and creation operators are defined by
$\hat b = \frac{1}{\sqrt{2\theta}} (\hat{x}+i\hat{y})$,
$\hat{b}^\dagger =\frac{1}{\sqrt{2\theta}} (\hat{x}-i\hat{y})$
and satisfy the Fock algebra $[ \hat b , \hat{b}^\dagger ] = 1$. 
The noncommutative configuration space can therefore be viewed as a boson Fock space spanned by the 
eigenstate $|n\rangle$ of the operator $b^{\dagger}b$. We refer to it as the
classical configuration space ($\mathcal{H}_c $)
\begin{equation}
\mathcal{H}_c = \textrm{span}\{ |n\rangle= 
\frac{1}{\sqrt{n!}}(\hat{b}^\dagger)^n |0\rangle\}_{n=0}^{n=\infty}
\label{3}
\end{equation}
where the span is taken over the field of complex numbers.

Then one introduces the Hilbert space
of the noncommutative quantum system, which is taken to be:
\begin{equation}
\mathcal{H}_q = \left\{ \psi(\hat{x},\hat{y}): 
\psi(\hat{x},\hat{y})\in \mathcal{B}
\left(\mathcal{H}_c\right),\;
{\rm tr_c}(\psi^\dagger(\hat{x},\hat{y})
\psi(\hat{x},\hat{y})) < \infty \right\}.
\label{4}
\end{equation}
Here ${\rm tr_c}$ denotes the trace over noncommutative 
configuration space and $\mathcal{B}\left(\mathcal{H}_c\right)$ 
the set of bounded operators on $\mathcal{H}_c$. 
This space has a natural inner product and norm 
\begin{equation}
\left(\phi(\hat{x}, \hat{y}), \psi(\hat{x},\hat{y})\right) = 
{\rm tr_c}(\phi(\hat{x}, \hat{x})^\dagger\psi(\hat{x}, \hat{y}))
\label{inner}
\end{equation}
and forms a Hilbert space \cite{hol}. An important notation that we adopt is the following. States in the noncommutative configuration space are denoted by $|\cdot\rangle$ and 
states in the quantum Hilbert space by $\psi(\hat{x},\hat{y})\equiv |\psi)$ to distinguish between them.  
A unitary representation of the noncommutative Heisenberg algebra 
in terms of operators $\hat{X}$, $\hat{Y}$, $\hat{P}_x$ and $\hat{P}_y$ acting on the states of the quantum Hilbert space 
(\ref{4}) (assuming commutative momenta) is easily found to be 
\begin{eqnarray}
\hat{X}\psi(\hat{x},\hat{y}) &=& \hat{x}\psi(\hat{x},\hat{y})\quad,\quad
\hat{Y}\psi(\hat{x},\hat{y}) = \hat{y}\psi(\hat{x},\hat{y})\nonumber\\
\hat{P}_x\psi(\hat{x},\hat{y}) &=& \frac{\hbar}{\theta}[\hat{y},\psi(\hat{x},\hat{y})]\quad,\quad
\hat{P}_y\psi(\hat{x},\hat{y}) = -\frac{\hbar}{\theta}[\hat{x},\psi(\hat{x},\hat{y})]~;~\hat{P}=\hat{P}_x +i\hat{P}_y.
\label{action}
\end{eqnarray}
The minimal uncertainty states on noncommutative 
configuration space (isomorphic to boson Fock space), 
are well known to be the normalized coherent states \cite{klaud}
\begin{equation}
\label{cs} 
|z\rangle = e^{-z\bar{z}/2}e^{z b^{\dagger}} |0\rangle
\end{equation}
where, $z=\frac{1}{\sqrt{2\theta}}\left(x+iy\right)$ 
is a dimensionless complex number. These states provide an overcomplete 
basis on the noncommutative configuration space. 
Corresponding to these states, a state 
(operator) in quantum Hilbert space can be constructed as follows
\begin{equation}
|z, \bar{z} )=\frac{1}{\sqrt{\theta}}|z\rangle\langle z|
\label{csqh}
\end{equation}
which satisfy 
\begin{equation}
\hat B|z, \bar{z})=z|z, \bar{z})~;~\hat B=\frac{1}{\sqrt{2\theta}}(\hat{X}+i\hat{Y})~,~[\hat B, \hat B^{\ddagger}]=1.
\label{p1}
\end{equation}
Writing the trace in terms of coherent states (\ref{cs}) and using 
$|\langle z|w\rangle|^2=e^{-|z-w|^2}$ it is easy to see that 
\begin{equation}
(z, \bar{z}|w, \bar{w})=\frac{1}{\theta}tr_{c}
(|z\rangle\langle z|w\rangle\langle w|)=
\frac{1}{\theta}|\langle z|w\rangle|^2=\frac{1}{\theta}e^{-|z-w|^2}
\label{p2}
\end{equation}
which shows that $|z, \bar{z})$ is indeed a Hilbert-Schmidt operator.  
The `position' representation of a state 
$|\psi)=\psi(\hat{x},\hat{y})$ can now be constructed as
\begin{equation}
(z, \bar{z}|\psi)=\frac{1}{\sqrt\theta}tr_{c}
(|z\rangle\langle z| \psi(\hat{x},\hat{y}))=
\frac{1}{\sqrt\theta}\langle z|\psi(\hat{x},\hat{y})|z\rangle.
\label{posrep}
\end{equation}
We now introduce the momentum eigenstates normalised such that $(p'|p)=\delta(p-p')=\delta(p_x-p'_x)\delta(p_y-p'_y)$
\begin{eqnarray}
|p)=\sqrt{\frac{\theta}{2\pi\hbar^{2}}}e^{i\sqrt{\frac{\theta}{2\hbar^2}}
(\bar{p}b+pb^\dagger)}~;~\hat{P}_i |p)=p_i |p)~,~p=p_x +ip_y
\label{eg}
\end{eqnarray}
which satisfy the completeness relation
\begin{eqnarray}
\int d^{2}p~|p)(p|=1_{q}~.
\label{eg5}
\end{eqnarray}
We now observe that the wave-function of a ``free particle" on the noncommutative plane is given by \cite{sgfgs}
\begin{eqnarray}
(z, \bar{z}|p)=\frac{1}{\sqrt{2\pi\hbar^{2}}}
e^{-\frac{\theta}{4\hbar^{2}}\bar{p}p}
e^{i\sqrt{\frac{\theta}{2\hbar^{2}}}(p\bar{z}+\bar{p}z)}~.
\label{eg3}
\end{eqnarray}
The completeness relations for the position eigenstates $|z,\bar{z})$ (which is an important
ingredient in the construction of the path integral representation) reads
\begin{eqnarray}
\int \frac{\theta dzd\bar{z}}{2\pi}~|z, \bar{z})\star(z, \bar{z}|=1_{q}
\label{eg6}
\end{eqnarray}
where the star product between two functions 
$f(z, \bar{z})$ and $g(z, \bar{z})$ is defined as
\begin{eqnarray}
f(z, \bar{z})\star g(z, \bar{z})=f(z, \bar{z})
e^{\stackrel{\leftarrow}{\partial_{\bar{z}}}
\stackrel{\rightarrow}{\partial_z}} g(z, \bar{z})~.
\label{eg7}
\end{eqnarray}
This can be proved by using (\ref{eg3}) and computing
\begin{eqnarray}
\int \frac{\theta dzd\bar{z}}{2\pi}
(p'|z, \bar{z})\star(z, \bar{z}|p)=
e^{-\frac{\theta}{4\hbar^{2}}(\bar{p}p+\bar{p}'p')}
e^{\frac{\theta}{2\hbar^{2}}\bar{p}p'}\delta(p-p')=(p'|p)~.
\label{eg8}
\end{eqnarray}
Thus, the position representation of the noncommutative system maps quite naturally to the Voros plane.

It is possible to decompose the star product $\star=e^{\stackrel{\leftarrow}{\partial_{\bar{z}}}
\stackrel{\rightarrow}{\partial_z}}$ by introducing a further degree of freedom (apart from constant factors)
\begin{eqnarray}
1_Q &=&\int dzd\bar{z}~|z, \bar{z})\star(z, \bar{z}|\nonumber\\
&=&\int dzd\bar{z}\int dvd\bar{v}~e^{-|v|^2}|z)e^{\bar{v}\stackrel{\leftarrow}{\partial_{\bar{z}}}+
v\stackrel{\rightarrow}{\partial_z}}(z|\nonumber\\
&=&\int dzd\bar{z}\int dvd\bar{v}~|z; v)(z; v|.
\label{eg8a1}
\end{eqnarray}
The states $|z; v)$ introduced above have the following additional properties
\begin{eqnarray}
|z; v)&=&e^{-\bar{v}v/2}e^{\bar{v}\partial_{\bar z}}|z)\nonumber\\
&=&e^{\frac{1}{2}(\bar{z}v-\bar{v}z)}|z\rangle \langle z+v|~;~ z, v \in \mathcal{C}
\label{eg8a2}
\end{eqnarray}
and 
\begin{eqnarray}
\hat{B}|z; v)=z|z; v)~;~\forall~ v.
\label{eg8a3}
\end{eqnarray}
For $v=0$, $|z; v)$ reduces to $|z, \bar z)$ given in eq.(\ref{csqh}).

\noindent The overlap of the $|z; v)$ states with the momentum eigenstates $|p)$ reads
\begin{eqnarray}
(z; v|p)=\frac{1}{\sqrt{2\pi\hbar^{2}}}
e^{-\frac{\theta}{4\hbar^{2}}\bar{p}p}
e^{\frac{i}{\hbar}\sqrt{\frac{\theta}{2}}[p\bar{z}+\bar{p}(z+v)]}e^{-\frac{1}{2}\bar{v}v}~.
\label{eg3z}
\end{eqnarray}
We introduce one further notational convention which we shall require later. For any operator $\hat{O}$ acting on the quantum Hilbert space, we may define left and right action (denoted by subscripted $L$ and $R$) as
follows:
\begin{eqnarray}
\hat{O}_{L}\psi=\hat{O}\psi~,~\hat{O}_{R}\psi=\psi\hat{O}~;~\forall ~\psi\in\mathcal{H}_{q}.
\label{n1}
\end{eqnarray}
In this language, for instance, the complex momenta $\hat{P}$ may be written as
\begin{eqnarray}
\hat{P}=i\hbar\sqrt{\frac{2}{\theta}}[\hat{B}_{R}-\hat{B}_{L}]~,~\hat{P}^{\ddagger}=i\hbar\sqrt{\frac{2}{\theta}}[\hat{B}^{\ddagger}_{L}-\hat{B}^{\ddagger}_{R}].
\label{n2}
\end{eqnarray}
In this way we have
\begin{eqnarray}
(z; v|\hat{B}^{\ddagger}_{L}|\psi)&=&e^{\frac{1}{2}(\bar{v}z-\bar{z}v)}\langle z|b^{\dagger}\psi|z+v\rangle
=\bar{z}(z; v|\psi),\nonumber\\
(z; v|\hat{B}_{R}|\psi)&=&e^{\frac{1}{2}(\bar{v}z-\bar{z}v)}\langle z|\psi b|z+v\rangle
=(z+v)(z; v|\psi),\nonumber\\
(z; v|\hat{B}_{L}|\psi)&=&e^{\frac{1}{2}(\bar{v}z-\bar{z}v)}\langle z|b\psi|z+v\rangle,\nonumber\\
(z; v|\hat{B}^{\ddagger}_{R}|\psi)&=&e^{\frac{1}{2}(\bar{v}z-\bar{z}v)}\langle z|\psi b^{\dagger}|z+v\rangle.
\label{n3}
\end{eqnarray}
Also, it is to be noted that the left and right operators commute with each other, 
since $\hat{A}_{L}\hat{B}_{R}\psi=\hat{A}\psi\hat{B}=\hat{B}_{R}\hat{A}_{L}\psi$.

With the above formalism and the completeness relations for the
momentum and the position eigenstates 
(\ref{eg5}, \ref{eg6}) in place, we now proceed to write down the
path integral for the propagation kernel
on the two dimensional noncommutative plane. This reads (apart from a constant factor)
\begin{eqnarray}
(z_f; v_f, t_f|z_0;v_0,  t_0)&=&\lim_{n\rightarrow\infty}\int
\prod_{j=1}^{n}(d\bar{\mu}_{j}d\mu_{j})~(z_f;v_f, t_f|z_n;v_n, t_n)
(z_n;v_n, t_n|....|z_1; v_1, t_1)(z_1; v_1, t_1|z_0; v_0, t_0)~.
\label{pint1}
\end{eqnarray}
The Hamiltonian (acting on the quantum Hilbert space) for a particle in the presence of a potential on the noncommutative plane reads
\begin{eqnarray}
\hat{H}=\frac{\vec{P}^2}{2m}+:V(B^{\dagger},B):\,.
\label{hamil}
\end{eqnarray} 
With this Hamiltonian, we now compute the propagator over a small segment in the
above path integral (\ref{pint1}). With the help of (\ref{eg5}) and (\ref{eg3}), we have
\begin{eqnarray}
(z_{j+1}; v_{j+1}, t_{j+1}|z_j; v_j, t_j)&=&(z_{j+1}; v_{j+1} |e^{-\frac{i}{\hbar}\hat{H}\tau}|z_j;  v_j)\nonumber\\
&=&(z_{j+1}; v_{j+1}|1-\frac{i}{\hbar}\hat{H}\tau +O(\tau^2)|z_j; v_{j})\nonumber\\
&=&\int_{-\infty}^{+\infty}d^{2}p_j~e^{-\frac{\theta}{2\hbar^{2}}\bar{p}_j p_{j}}
e^{\frac{i}{\hbar}\sqrt{\frac{\theta}{2}}\left[p_{j}(\bar{z}_{j+1}-\bar{z}_{j})+\bar{p}_{j}(z_{j+1}-z_{j})\right]}\nonumber\\
&&\times e^{-\frac{i}{\hbar}\tau[\frac{\bar{p}_j p_{j}}{2m}+V(\bar{z}_{j+1}, z_{j})]}
e^{-\frac{1}{2}(|v_{j+1}|^2 +|v_j|^2)}e^{\frac{i}{\hbar}\sqrt{\frac{\theta}{2}}\left(\bar{p}_{j}v_{j+1}-p_{j}\bar{v}_{j}\right)}+O(\tau^2)~.
\label{pint2}
\end{eqnarray} 
Substituting the above expression in eq.(\ref{pint1}), we obtain (apart from a constant factor)
\begin{eqnarray}
(z_f; v_f, t_f|z_0;v_0,  t_0)&=&\lim_{n\rightarrow\infty}\int 
\prod_{j=1}^{n}(d\bar{\mu}_{j}d\mu_{j})\prod_{j=0}^{n}d^{2}p_{j}\nonumber\\
&&\exp\sum_{j=0}^{n}\left[\frac{i}{\hbar}\sqrt{\frac{\theta}{2}}\left[p_{j}\left\{\bar{z}_{j+1}-\bar{z}_{j}\right\}+\bar{p}_{j}\left\{z_{j+1}-z_{j}\right\}\right] 
+\alpha \bar{p}_{j}p_{j}
-\frac{i}{\hbar}\tau V(\bar{z}_{j+1},z_{j})\right.\nonumber\\
&&\left.~~~~~~~~~~~~~~-\frac{1}{2}(|v_{j+1}|^2 +|v_j|^2)+\frac{i}{\hbar}\sqrt{\frac{\theta}{2}}\left(\bar{p}_{j}v_{j+1}-p_{j}\bar{v}_{j}\right)\right]
\label{pint3}
\end{eqnarray} 
where $\alpha=-(\frac{i\tau}{2m\hbar}+\frac{\theta}{2\hbar^{2}})$.

\noindent One can now carry out the momentum integral to obtain
\begin{eqnarray}
(z_f; v_f, t_f|z_0; v_0, t_0)&=&\lim_{n\rightarrow\infty}N\int\prod_{j=1}^{n}(d\bar{\mu}_{j}d\mu_{j})\nonumber\\
&&\exp\sum_{j=0}^{n}\left[\frac{\theta}{2\hbar^2 \alpha}[|z_{j+1}-z_j|^2 -v_{j+1}\bar{v}_{j}+(\bar{z}_{j+1}-\bar{z}_{j})v_{j+1}-(z_{j+1}-z_{j})\bar{v}_{j}]
-\frac{i}{\hbar}\tau V(\bar{z}_{j+1},z_{j})\right.\nonumber\\
&&\left.~~~~~~~~~~~~~~~ -\frac{1}{2}(|v_{j+1}|^2 +|v_j|^2)\right].
\label{pintegral1}
\end{eqnarray} 
Now using the fact that $z_{j}=z(j\tau)$ and $z_{j+1}-z_{j}=\tau\dot{z}(j\tau)+O(\tau^{2})$ and taking the limit $\tau\rightarrow 0$, we finally arrive at the path integral representation of the propagator
\begin{eqnarray}
(z_f; v_f, t_f|z_0;v_0, t_0)=N\int \mathcal{D}\bar{\mu}\mathcal{D}\mu~
\exp(\frac{i}{\hbar}S)
\label{pintegral3}
\end{eqnarray} 
where $S$ is the action given by
\begin{eqnarray}
S=\int_{t_{0}}^{t_{f}}dt \left[-\frac{\hbar^2}{m\theta}\bar{v}v+i\hbar\left\{\dot{\bar z}v-\dot{z}\bar{v}
+\frac{1}{2}(\dot{\bar v}v-\bar{v}\dot{v})\right\}- V(\bar{z}(t),z(t))\right]~.
\label{action_ncqm}
\end{eqnarray} 
Before proceeding further, we would like to point out that this form of the action looks very similar in structure
to the action involving the auxiliary field considered in \cite{sgfgs}. This can be realized by recasting the above action
in the following form
\begin{eqnarray}
S=\int_{t_{0}}^{t_{f}}dt \left[\bar{v}\left(-i\hbar\partial_{t}-\frac{\hbar^2}{m\theta}\right)v
+i\hbar(\dot{\bar z}v-\dot{z}\bar{v})\right].
\label{ac_ncq}
\end{eqnarray} 
The above action can be readily identified with  the action involving the auxiliary field \cite{sgfgs}. As a consistency check,
we now quantize this theory to see whether one recovers the noncommutative Heisenberg algebra. To do
this we first set $v=v_1 +iv_2$ and $z=\frac{1}{\sqrt{2\theta}}(x+iy)$ to find that this is a constrained system with 
the following second class constraints
\begin{eqnarray}
\Lambda_1&=&p_{v_1}+\hbar v_2\approx0,\nonumber\\
\Lambda_2&=&p_{v_2}-\hbar v_1\approx0,\nonumber\\
\Lambda_3&=&p_{x}+\sqrt{\frac{2}{\theta}}\hbar v_2\approx0,\nonumber\\
\Lambda_4&=&p_{y}-\sqrt{\frac{2}{\theta}}\hbar v_1\approx0,
\label{secclass}
\end{eqnarray} 
where the symbol $\approx0$ signifies weakly zero. Computing the Dirac brackets \cite{dirac} yield 
\begin{eqnarray}
\{x, y\}_{DB}=\frac{\theta}{\hbar}~,~\{x, p_x\}_{DB}=1=\{x, p_x\}_{DB}
~,~\{x, v_2\}_{DB}=-\frac{1}{\hbar}\sqrt{\frac{\theta}{2}}~.
\label{dirac}
\end{eqnarray} 
Replacing $\{., .\}_{DB}\rightarrow\frac{1}{i\hbar}[., .]$ indeed yields the noncommutative
Heisenberg algebra.

The connection of the above propagator with the propagator in the $|z)$-basis (which
has been derived in \cite{sgfgs}) can be shown in the following way. We note that the propagator in the $|z)$-basis is related
to the propagator in the $|z; v)$-basis as
\begin{eqnarray}
(z_f, t_f|z_0, t_0)=N e^{-\vec{\partial}_{z_f}\vec{\partial}_{\bar{z}_0}}\int d\bar{v}dv~(z_f; v, t_f|z_0; v, t_0).
\label{pr1}
\end{eqnarray} 
Setting $v_f =v_0 =v$ in eq.(\ref{pintegral1}) and integrating over $v$, $\bar v$ leads to
\begin{eqnarray}
\int d\bar{v}dv~(z_f; v, t_f|z_0; v, t_0)=N\int \mathcal{D}\bar{z}\mathcal{D}z~\exp\left\{\frac{i}{\hbar}\int_{t_0}^{t_f}\theta m \dot{\bar z}(t)\left(1+\frac{im\theta}{\hbar}\partial_t\right)^{-1}\dot{z}(t)\right\}.
\label{pr2}
\end{eqnarray} 
From eq(s)(\ref{pr1}, \ref{pr2}), we get
\begin{eqnarray}
(z_f, t_f|z_0, t_0)=N e^{-\vec{\partial}_{z_f}\vec{\partial}_{\bar{z}_0}}\int \mathcal{D}\bar{z}\mathcal{D}z~\exp\left\{\frac{i}{\hbar}\int_{t_0}^{t_f}\theta m \dot{\bar z}(t)\left(1+\frac{im\theta}{\hbar}\partial_t\right)^{-1}\dot{z}(t)\right\}
\label{pr3}
\end{eqnarray} 
which is the propagator in \cite{sgfgs}. For the sake of completeness, we compute the free particle propagator in the
$|z; v)$-basis. This reads
\begin{eqnarray}
(z_f; v_f, t_f|z_0;v_0, t_0)=\frac{\theta m}{(i\hbar T +\theta m)}\exp\left\{-\frac{(|v_f|^2 +|v_0|^2)}{2}
-\frac{\theta m}{(iT +\theta m)}[|z_f -z_0|^2 -\bar{v}_{0}(z_f -z_0) +v_{f}(\bar{z}_f -\bar{z}_0)
-\bar{v}_{0}v_{f} ]\right\}\nonumber\\
\label{prop1}
\end{eqnarray}
where $T=t_f -t_0$. To obtain the free particle propagator in the $|z)$-basis, we first set $v_f =v_0 =v$ and then use eq.(\ref{pr1}) to get
\begin{eqnarray}
(z_f, t_f|z_0, t_0)=\frac{\theta m}{(i\hbar T +\theta m)}\exp\left\{-\frac{\theta m}{(iT +\theta m)}|z_f -z_0|^2\right\}.\nonumber\\
\label{prop2}
\end{eqnarray} 
This is precisely the result obtained in \cite{sgfgs} and hence verifies the consistency of eq.(\ref{pr1}).

We now move on to demonstrate that the path integral formulation developed in \cite{sgfgs} leads to a Chern-Simons 
quantum mechanics. To establish this link, we first write down the path integral representation of the propagator 
in the $|z)$-basis \cite{sgfgs}
\begin{eqnarray}
(z_f, t_f|z_0, t_0)=&&\lim_{n\rightarrow\infty}\int\prod_{j=1}^{n}(d\bar{z}_{j}dz_{j})\prod_{j=0}^{n}d^{2}p_{j} e^{-\vec{\partial}_{z_f}\vec{\partial}_{\bar{z}_0}}\nonumber\\
&&\times\exp\sum_{j=0}^{n}\left[\frac{i}{\hbar}\sqrt{\frac{\theta}{2}}\left[p_{j}\left\{\bar{z}_{j+1}-\bar{z}_{j}\right\}+\bar{p}_{j}\left\{z_{j+1}-z_{j}\right\}\right] 
+\alpha \bar{p}_{j}p_{j}+\frac{\theta}{2\hbar^2}p_{j+1}\bar{p}_{j}
-\frac{i}{\hbar}\tau V(\bar{z}_{j+1},z_{j})\right].
\label{pr4}
\end{eqnarray} 
where $\alpha=-(\frac{i\tau}{2m\hbar}+\frac{\theta}{2\hbar^2})$. Using this value of $\alpha$, it is easy to see that
\begin{eqnarray}
\alpha \bar{p}_{j}p_{j}+\frac{\theta}{2\hbar^2}p_{j+1}\bar{p}_{j}=-\frac{i\tau}{2m\hbar}\bar{p}_{j}p_{j}
+\frac{\theta}{2\hbar^2}\bar{p}_{j}(p_{j+1}-p_j).
\label{pr5}
\end{eqnarray} 
Substituting the above relation in eq.(\ref{pr4}) and taking the $\tau\rightarrow 0$, we get
\begin{eqnarray}
(z_f, t_f|z_0,t_0)=N\int \mathcal{D}\bar{z}\mathcal{D}z \mathcal{D}\bar{p}\mathcal{D}p~
e^{-\vec{\partial}_{z_f}\vec{\partial}_{\bar{z}_0}}\exp(\frac{i}{\hbar}S)
\label{pintegral6}
\end{eqnarray} 
where $S$ is the action given by
\begin{eqnarray}
S=\int_{t_0}^{t_f}dt~\left[\sqrt{\frac{\theta}{2}}(p\dot{\bar z}+\bar{p}\dot{z})-\frac{\bar{p}p}{2m}-\frac{i\theta}{2\hbar}\bar{p}\dot{p}\right].
\label{pr7}
\end{eqnarray} 
The important point to note about this form of the action is the $\bar{p}\dot{p}$ term. This is a Chern-Simons term
in momentum space and is similar to the $\bar{v}\dot{v}$ term arising in eq.(\ref{action_ncqm}).
Note that this term is purely of noncommutative origin and vanishes in the $\theta\rightarrow0$ limit.
Such a term has been considered earlier in \cite{jianjing}, however, it comes out naturally in our path integral formalism.
It is also reassuring to note that imposing the second class constraints $\Lambda_3$ and $\Lambda_4$ strongly
(which implies $v=-\frac{i}{\hbar}\sqrt{\frac{\theta}{2}}p$) and substituting this in the action (\ref{action_ncqm})
yields eq.(\ref{pr7}).

The above analysis can also be carried out for a particle moving in a magnetic field in the noncommutative plane.
The path integral representation of the propagator in this case reads \cite{sgfgs1}
\begin{eqnarray}
(z_f, t_f|z_0, t_0)=&&\lim_{n\rightarrow\infty}\int\prod_{j=1}^{n}(d\bar{z}_{j}dz_{j})\prod_{j=0}^{n}d^{2}p_{j} e^{-\vec{\partial}_{z_f}\vec{\partial}_{\bar{z}_0}}\nonumber\\
&&\times\exp\sum_{j=0}^{n}\left[\frac{i}{\hbar}\sqrt{\frac{\theta}{2}}\left[p_{j}\left\{\bar{z}_{j+1}-\bar{z}_{j}\right\}+\bar{p}_{j}\left\{z_{j+1}-z_{j}\right\}\right] 
+\alpha \bar{p}_{j}p_{j}+\frac{\theta}{2\hbar^2}p_{j+1}\bar{p}_{j}
+\frac{eB}{2m\hbar}\sqrt{\frac{\theta}{2}}\tau(p_{j}\bar{z}_{j+1}-\bar{p}_{j}z_{j})\right].\nonumber\\
\label{jphys}
\end{eqnarray} 
Now using eq.(\ref{pr5}) and the fact that $z_{j}=z(j\tau)$ and $z_{j+1}=z_{j}+\tau\dot{z}(j\tau)+O(\tau^2)$
followed by the $\tau\rightarrow0$ limit leads to eq.(\ref{pintegral6}) with the following form of the action
\begin{eqnarray}
S&=&\int_{t_0}^{t_f}dt~\left[\sqrt{\frac{\theta}{2}}(p\dot{\bar z}+\bar{p}\dot{z})-\frac{\bar{p}p}{2m}-\frac{i\theta}{2\hbar}\bar{p}\dot{p}-\frac{ieB}{2m}\sqrt{\frac{\theta}{2}}(p\bar{z}-\bar{p}z)\right]\nonumber\\
&=&\int_{t_0}^{t_f}dt~\left[(p_{i}+eA_{i})\dot{x}_{i}+\frac{\theta}{2\hbar}\epsilon_{ij}p_{i}\dot{p}_{j}
-\frac{p_{i}^2}{2m}\right].
\label{jphys1}
\end{eqnarray} 
where $A_{i}=-\frac{B}{2}\epsilon_{ij}x_{j}, (i,j=1,2)$. This action derived
here from a path integral framework is precisely the one written down in \cite{jianjing}. 

\noindent It is also interesting to note that the origin of the Chern-Simons term in momentum can be understood
from the point of view of a Bopp-shift. This can be seen by writing down 
the action in the first order form (in commutative space)
\begin{eqnarray}
S=\int_{t_0}^{t_f}dt~(p_{i}\dot{x}_{i}-H).
\label{pr7l}
\end{eqnarray} 
Now replacing $x_{i}$ by the Bopp-shifted variable $\tilde{x}_{i}$ given by
\begin{eqnarray}
\tilde{x}_{i}=x_{i}-\frac{\theta}{2\hbar}\epsilon_{ij}p_{j}~;~\epsilon_{ij}=-\epsilon_{ji}~,~(i,j=1,2)
\label{boppshift}
\end{eqnarray} 
we get (with $H=\frac{p_{i}^2}{2m}$)
\begin{eqnarray}
S\rightarrow\tilde{S}&=&\int_{t_0}^{t_f}dt~\left(p_{i}\dot{\tilde{x}}_{i}-\frac{p_{i}^2}{2m}\right)\nonumber\\
&=&\int_{t_0}^{t_f}dt~\left(p_{i}\dot{x}_{i}-\frac{p_{i}^2}{2m}-\frac{\theta}{2\hbar}\epsilon_{ij}p_{i}\dot{p}_{j}\right).
\label{bs1act}
\end{eqnarray} 
This expression clearly shows that the action $S$ gets augmented by the Chern-Simons term in momentum (which is 
of noncommutative origin) when the variable $x_{i}$ is Bopp-shifted.

The action (\ref{pr7}) can also be readily generalized to the relativistic case and reads
\begin{eqnarray}
S=\int_{t_0}^{t_f}d\tau~\left[p_{\mu}\dot{x}^{\mu}-\lambda(\tau)(p_{\mu}p^{\mu}+m^2)+\frac{\theta^{\mu\nu}}{2\hbar}p_{\mu}\dot{p}_{\nu}\right]
\label{pr7v}
\end{eqnarray} 
where $\tau$ is the proper time and $\lambda(\tau)$ is the Lagrange multiplier enforcing the constraint $p_{\mu}p^{\mu}+m^2=0$. This form of the action has been considered earlier in \cite{deriglazov} but 
arises here as a natural generalization of the non-relativistic form of the action (\ref{pr7}) derived from a path integral approach. In \cite{spal}, a relativistic generalization of the action following from the non-relativistic form of the path integral
was presented, however, the noncommutative term differed from the Chern-Simons form and appeared to be quadratic in the momentum.
%%%%%%%%%%%%%%%%%%%%%%%%%%%%%%%%%%%%%%%%%
%%%%%%%%%%%%%%%%%%%%%%%%%%%%%%%%%%%%%%%%%

%%%%%%%%%%%%  Map  %%%%%%%%%%%%%%%%%%%%%%%%%%

We now proceed to show that the problem of  a quantum Hall system in commutative space can be mapped
to the problem of a particle in a noncommutative plane in the presence of a magnetic field. To proceed, we first
write down the Hamiltonian of the quantum Hall system in the presence of a harmonic oscillator potential
\begin{eqnarray}
\hat{H}=2\hbar\omega_{c}a_{-}^{\dagger}a_{-}+\frac{k}{2}(\hat{X}^2 +\hat{Y}^2)~;~k=m\omega^2~,~\omega_{c}=\frac{eB}{2m}
\label{qh1}
\end{eqnarray} 
where $\omega_{c}$ is the cyclotron frequency and
\begin{eqnarray}
\hat{X}&=&\sqrt{\frac{\hbar}{2m\omega_c}}(a_x +a_{x}^{\dagger})~;
~\hat{Y}=\sqrt{\frac{\hbar}{2m\omega_c}}(a_y +a_{y}^{\dagger})\nonumber\\
a_{\pm}&=&\frac{1}{\sqrt{2}}(a_x +a_y).
\label{qh1a}
\end{eqnarray} 
Now eq.(\ref{n3}) points out that the operators $a_{-}$ and  $a_{+}$ and their hermitian conjugates 
can be mapped to the noncommutative operators in the following way
\begin{eqnarray}
a_{-}\leftrightarrow B^{\ddagger}_{R}~,~a_{+}\leftrightarrow B_{L}\nonumber\\
a_{-}^{\dagger}\leftrightarrow B_{R}~,~a_{+}^{\dagger}\leftrightarrow B_{L}^{\ddagger}
\label{qh2}
\end{eqnarray} 
since the additional degree of freedom $v$ can be interpreted as a coherent state label in the space of Landau levels in a quantum Hall system \footnote{Note that the coordinate $z$ with the knowledge of the additional degree of freedom
$v$ completely specifies the state of the particle}. The above mapping of operators can be justified from the fact that
the right action operators $B_{R}, B_{R}^{\ddag}$ act on the additional degrees of freedom.
The map between the operators can be used to map the commutative problem (\ref{qh1}) to a noncommutative problem.
The Hamiltonian of the corresponding noncommutative problem reads
\begin{eqnarray}
\hat{H}=2\hbar\omega_{c}+2\tilde{k}( B_{L}^{\ddagger} B_{L}+ B_{R} B_{R}^{\ddagger}+B_{L} B_{R}^{\ddagger}
+B_{L}^{\ddagger}B_{R})~;~\tilde{k}=\frac{\hbar\omega^2}{4\omega_c}~.
\label{qh3}
\end{eqnarray} 
The action for this Hamiltonian can be easily computed from the result \cite{rohwer}
\begin{eqnarray}
S=\int_{t_{0}}^{t_{f}} dt~(z; v|i\hbar\partial_{t}-\hat{H}|z; v).
\label{qh4}
\end{eqnarray} 
Substituting the form of the Hamiltonian in the above action and using eq.(\ref{n3}), we get
\begin{eqnarray}
S=\int_{t_{0}}^{t_{f}}dt \left[-2(\hbar\omega_c +\tilde{k})\bar{v}v-i\hbar\bar{v}\dot{v}
+i\hbar\left\{\dot{\bar z}-2(\hbar\omega_c +2\tilde{k})\bar{z}\right\}v- i\hbar\left\{\dot{z}-2(\hbar\omega_c +2\tilde{k})z\right\}\bar{v}-2(\hbar\omega_c +4\tilde{k})\bar{z}z\right]~.
\label{qh5}
\end{eqnarray} 
Putting the above form of the action in eq.(\ref{pintegral3}) and integrating the additional degrees 
of freedom using eq.(\ref{action_ncqm}), we obtain the following form of the action 
in the $|z)$ coherent state representation
\begin{eqnarray}
S=\int_{t_{0}}^{t_{f}} dt\left[\frac{\hbar^2}{2(\hbar\omega_c +\tilde{k})}
\left(\partial_{t}+\frac{2i}{\hbar}[\hbar\omega_{c}+2\tilde{k}]\right)\bar{z}
\left(1+\frac{i\hbar}{2(\hbar\omega_c +\tilde{k})}\partial_{t}\right)^{-1}
\left(\partial_{t}-\frac{2i}{\hbar}[\hbar\omega_{c}+2\tilde{k}]\right)z -2(\hbar\omega_c +4\tilde{k})\bar{z}z\right].
\label{qh6}
\end{eqnarray} 
This form of the action can be recast to the action for a particle in a magnetic field in a harmonic oscillator
potential  \cite{sgfgs1}
\begin{eqnarray}
S=\int_{t_{0}}^{t_{f}} dt\left[\theta \tilde{m}
\left(\partial_{t}+\frac{ie\tilde{B}}{2\tilde{m}}\right)\bar{z}
\left(1+\frac{i\theta\tilde m}{\hbar}\partial_{t}\right)^{-1}
\left(\partial_{t}+\frac{ie\tilde B}{2\tilde m}\right)z 
-\left(\frac{e^2 \tilde{B}^2}{4\tilde{m}}+\tilde{m}\tilde{\omega}^2\right)\bar{z}z\right]
\label{qh7}
\end{eqnarray} 
with the following identifications
\begin{eqnarray}
\theta \tilde{m}&=&\frac{\hbar^2}{2(\hbar\omega_c +\tilde{k})},\nonumber\\
\frac{e\tilde{B}}{2\tilde{m}}&=&\frac{2}{\hbar}(\hbar\omega_c +2\tilde{k}),\nonumber\\
\left(\frac{e^2 \tilde{B}^2 }{4\tilde{m}}+\tilde{m}\tilde{\omega}^2\right)\theta&=&2(\hbar\omega_c +4\tilde{k}).
\label{qh8}
\end{eqnarray} 
The above set of equations imply
\begin{eqnarray}
\tilde{\omega}^2=\frac{4\tilde{k}\omega_{c}}{\hbar}~.
\label{qh9}
\end{eqnarray} 
We would like to mention that similar identifications have also been noted earlier in the literature
\cite{duval, nair, sg, sgbc} where they were found as rescalings 
from one set of parameters to another set of parameters.
Setting $\tilde{m}=m$ leads to the following value of the noncommutative parameter $\theta$:
\begin{eqnarray}
\theta&=&\frac{\hbar^2}{2m(\hbar\omega_{c}+\tilde{k})}\nonumber\\
&=&\frac{\hbar}{eB}\left(1-\frac{\omega^2}{4\omega_{c}^{2}}+\mathcal{O}(\omega^4/\omega_{c}^4)\right)
\label{qh10}
\end{eqnarray} 
where we have used the fact that $\omega<<\omega_{c}$ in making a binomial expansion in the second line. This result has the same form as the result for the commutator of the relative coordinates (projected to the lowest Landau level) in a magnetic field in the presence of a harmonic oscillator potential computed in \cite{sggov}. This gives a clear understanding of the noncommutativity in the lowest Landau level of a quantum Hall system from a path integral point of view.

We now summarize our findings. In this paper, using the recently proposed formulation of quantum mechanics on noncommutative plane, we derive the path integral representation of the propagation amplitude for a particle using coherent states having additional degrees of freedom. We show the relation of this amplitude with the propagator computed using
coherent states without any additional degrees of freedom. We then demonstrate that the path integral formulation using coherent states on the noncommutative plane gives rise to a noncommutative Chern-Simons quantum mechanics. This
is shown by deriving a phase-space representation of the path integral. This observation is new and is the main result in this paper. We also establish the connection of this Chern-Simons term in momentum (which is of noncommutative origin)
with the Bopp-shift. Finally, we construct a map from the commutative quantum Hall system to a particle in a noncommutative plane moving in a magnetic field. This enables us to compute the value of the noncommutative parameter and we find that it agrees with the result  for the commutator of the relative coordinates (projected to the lowest Landau level) in a magnetic field in the presence of a harmonic oscillator potential. As a future prospect, it would be interesting to investigate possible connections of our work with the fractional quantum Hall effect \cite{haldane}.

%%%%%%%%%%%%%%%%%%%%%%%%%%%%%%%%%%%%%

%%%%%%%%%%%%%%%%%%%%%%%%%%%%%%%%%
%%%%%%%%%%%%%%%%%%%%%%%%%%%%%%%%%

%%%%%%%%%%%%%%%%%%%%%%%%%%%%%%%%%%%

%%%%%%%%%%%%%%%%%%%%%%%%%%%%%%%%%%%%%%%%%%%%%%%%%%%%%%%%%%%%

\vskip 1 cm

\noindent {\bf{Acknowledgements}} : This work was supported under a grant of the National Research Foundation of South Africa. 

%%%%%%%%%%%%%%%%%%%%%%%%%%%%%%%%%%%%%%%%%%%%%%%%%%%%%%%%%%%%%%%%%%%%%%%%%%%

%%%%%%%%%%%%%%%%%%%%%%%%%%%%%%%%%%%%%%%%%%%%%%%%%%%%%%%%%%%%%%%%%%%%%%%%%%%%
%%%%%%%%%%%%%%%%%%%%%%%%%%%%%%%%%%%%%%%%%%%%%%%%%%%%%%%%%%%%%%%%%%%%%%%%%%%%
\end{document}